\documentclass{article}

\usepackage[english]{babel}

\usepackage[letterpaper,top=2cm,bottom=2cm,left=2cm,right=2cm,marginparwidth=1.75cm]{geometry}

\usepackage{mathrsfs}
\usepackage{amsmath}
\usepackage{amssymb}
\usepackage{epsfig}
\usepackage{graphicx}
\usepackage{subfigure}
\usepackage{multirow}
\usepackage{epstopdf}
\usepackage{authblk}
\usepackage{cite}

\def\be{\begin{eqnarray}}
\def\en{\end{eqnarray}}
\def\bea{\begin{aligned}}
\def\ena{\end{aligned}}

\title{Quasi-two-body decays $B_{(s)}\to K^*\gamma\to K\pi\gamma$ in perturbative QCD approach}
\author{Zhi-Qing Zhang
\footnote{Electronic address: zhangzhiqing@haut.edu.cn}, Yan-Chao Zhao, Zhi-Lin Guan, Zhi-Jie Sun, Zi-Yu Zhang, Ke-Yi He} 
\affil{ Department of Physics, Henan University of Technology, Zhengzhou, Henan 450052, China } 

\begin{document}
\maketitle

\begin{abstract}
In this work we study the quasi-two-body decays $B\to K^*\gamma\to K\pi\gamma$ in the perturbative QCD (PQCD) approach.
The two-meson distribution amplitudes (DAs) are introduced to describe the final state interactions of the $K\pi$ pair,
which involve the time-like form factors and the Gegenbauer polynomials.
We calculate the CP averaged branching ratios for the  decays $B_{(s)}\to K^*\gamma\to K\pi\gamma$.
Our results are in agreement with the new update data measured by Belle II, which suggests these quasi-two-body decays are more
appropriate to be analyzed in three-body framework than in the two-body one.
We also predict the direct CP-violation asymmetries for the considered decay modes and find that $A_{CP}(B_{u,d}\to K^*\gamma\to K\pi\gamma)$
is small and less than $1\%$ in magnitude, while $A_{CP}(B_{s}\to K^*\gamma\to K\pi\gamma)$ is larger and can arrive at a few percent.
Our predictions can be tested by the future B meson experiments.
\end{abstract}

{\centering\section{INTRODUCTION}\label{intro}}

A lot of experimental studies on
the three-body B meson decays\cite{shiyan1,shiyan2,shiyan3,shiyan4,shiyan5,shiyan6,shiyan7,shiyan8,shiyan9} have been performed in recent years.
This kind of decay is getting more and more attentions, which are caused by the following reasons: (1) Many new resonance states
are observed in the invariant mass distributions of the three-body decays, which are difficult to understand as a
common meson or baryon, and called as exotic states. People are puzzled by their inner structures and proposed many assumptions,
such as compact $qq\bar q\bar q$ tetraquark, $qqqq\bar q$ pentaquark, loosely bound hadronic molecule, glueball and hybrid state, etc.
(2) These decays involve much more complicated QCD dynamics compared with two-body cases,
which impose a serve challenge to the present theoretical frameworks.
The hard b-quark decay kernels in three-body decays contain two virtual gluons at leading order, it is difficult to directly
evaluate due to the enormous number of diagrams. (3) Large direct CP asymmetries in localized regions of the phase space for three-body decays
are observed in
experiments. The measured CP violation is just a number in two-body decays, while a distribution in the Dalitz plot for three-body decays,
where the sign and magnitude vary from region to region. In order to study these decays, many approaches based on the symmetry principles
and factorization theorems have been proposed. The symmetry principles incudes the U-spin \cite{sym1,sym4,sym5,sym6}, flavor $SU(3)$
symmetry \cite{sym2,sym22,sym3,sym33}, topological diagram amplitude (TAD) approach
\cite{tpa}, etc. The factorization theorems includes the QCD-improved factorization approach \cite{qcdf1,qcdf2,qcdf3,qcdf4,qcdf5,qcdf6} and
the PQCD approach\cite{pqcd1,pqcd2,pqcd3,pqcd4,pqcd5,pqcd6,pqcd8,pqcd9,pqcd10,pqcd11},  where it has been proposed
that the factorization theorem of three-body B decays is approximately valid when two particles move
collinearly and the bachelor particle recoils back. Based on the quasi-two-body-decay mechanism, the two-hadron distribution
amplitudes (DAs) are introduced into the PQCD approach, where the strong dynamics between the two final hadrons in the resonant regions are
included.

On the experimental side, the decays $B^{0,+}\to K^{*0,+}\gamma$ were investigated by Belle II recently \cite{cankao}, and got their branching ratios through
the different decay modes
\be\label{23}
\bea
Br\left(B^0\to K^{*0}[K^+\pi^-]\gamma\right) &=(4.5 \pm 0.3 \pm 0.2) \times 10^{-5},\\
Br\left(B^0\to K^{*0}[K^0\pi^0]\gamma\right) &=(4.4 \pm 0.9 \pm 0.6) \times 10^{-5}, \\
Br\left(B^+\to K^{*+}[K^+\pi^0]\gamma\right) &=(5.0 \pm 0.5 \pm 0.4) \times 10^{-5},\\
Br\left(B^+\to K^{*+}[K^0\pi^+]\gamma\right) &=(5.4 \pm 0.6 \pm 0.4) \times 10^{-5}.
\ena
\en
Certainly, the two-body decays $B^{0,+}\to K^{*0,+}\gamma$ have been studies by the different theories \cite{beneke,matsumori,ali}. Here we would like
to study these three-body radiative decays in quasi-two-body mechanism by using PQCD approach.
After introduced the new non-perturbative inputs, the two-meson distribution amplitudes,
the factorization formula for the three-body decay $B\rightarrow h_1h_2h_3$ can be written as\cite{cff1,cff2}
\be
\mathcal{M}=\Phi_{B} \otimes H \otimes \Phi_{h_{1} h_{2}} \otimes \Phi_{h_{3}},
\en
where $\Phi_B(\Phi_{h_3})$ denotes the $B(h_3)$ meson DAs, $\Phi_{h_{1} h_{2}}$ is  the $h_1h_2$ two-meson DA,
and $\otimes$ means the convolution in parton momenta. Then the hard kernel H for the b quark decay,
similar to the two-body case, starts with the diagrams of single hard gluon exchange.

This paper is organized as follows. In Sec. II, the kinematic variables for the B meson three-body radiative
decays are defined. The considered two-meson ($K\pi$) P -wave DAs are  parametrized, whose normalization form factors are assumed to
take the relativistic Breit-Wigner (RBW) model. Then Feynman diagrams and analytical expressions for these decays are given.
In Sec. III, the numerical results are presented and discussed, where
we would compare our predictions with other theoretical and experimental results.
The summary is presented in the final part.


{\centering\section{THE FRAMEWORK}\label{framework}}

We begin with the parametrization of the kinematic variables involved in the decay $B\to K^*\gamma\to K\pi\gamma$. In the rest frame of the B meson,
we define the B meson momentum $P_B$, the K meson momentum $P_1$, the $\pi$ meson momentum $P_2$, the $K^*$ meson momentum$P=P_1+P_2$
and the $\gamma$ momentum $P_3$ in the light-cone coordinates as
\be
\bea
P_B &= \frac{m_B}{\sqrt{2}}(1,1,0_T),P = \frac{m_B}{\sqrt{2}}(1,\eta,0_T),P_3 = \frac{m_B}{\sqrt{2}}(0,1-\eta,0_T),\\
P_1 &= \frac{m_B}{\sqrt{2}}(\zeta,(1-\zeta)\eta,P_{1T}),P_2 = \frac{m_B}{\sqrt{2}}((1-\zeta),\zeta\eta,P_{2T}),
\ena
\en
with the B meson mass $m_B$ and the variable $\eta = P^2/m_B^2 = \omega^2/m_B^2$, $\omega$ being the invariant mass of the $K\pi$ pair and $\zeta$ is the momentum fraction for the K meson.
The momenta of the light quarks in the B meson and the $K^*$ meson as $k_B$ and $k$ respectively
\be
k_B = (0,\frac{m_B}{\sqrt{2}}x_1,k_{1T}), \;\; k = (\frac{m_B}{\sqrt{2}}z,0,k_{2T}),
\en
where $x_1$and $z$ are the momentum fractions.

{\subsection{Distribution amplitudes}}

The P-wave $K\pi$ two-meson distribution amplitudes are defined as \cite{pqcd6}
\be
\Phi_{K\pi}^{T}(z, \zeta, \omega)=\frac{1}{\sqrt{2 N_{c}}}\left[\gamma_{5} \epsilon\hspace{-1.5truemm}/_{T}p\hspace{-1.5truemm}/\phi_{K\pi}^{T}\left(z, \omega^{2}\right)
+\omega \gamma_{5} \epsilon\hspace{-1.5truemm}/_{T} \phi_{K\pi}^{a}\left(z, \omega^{2}\right)
+i \omega \frac{\epsilon^{\mu \nu \rho \sigma} \gamma_{\mu} \epsilon_{T \nu} p_{\rho} n_{-\sigma}}{p \cdot n_{-}} \phi_{K\pi}^{v}
\left(z, \omega^{2}\right)\right] \sqrt{\zeta(1-\zeta)},
\en
with the functions\cite{das}
\be
\bea
\phi_{K\pi}^{T}(z,\omega^2) &=\frac{3F^\perp_{K\pi}(\omega^2)}{\sqrt{2N_c}}z(1-z)[1+a^\perp_{1K^*}3t+a^\perp_{2K^*}\frac{3}{2}(5t^2-1)],\\
\phi_{K\pi}^{a}(z,\omega^2) &=\frac{3F^\parallel_{K\pi}(\omega^2)}{4\sqrt{2N_c}}t,\\
\phi_{K\pi}^{v}(z,\omega^2) &=\frac{3F^\parallel_{K\pi}(\omega^2)}{8\sqrt{2N_c}}(1+t^2),
\ena
\en
where $t = (1-2z)$ and the Gegenbauer moments associated with transverse polarization
$a^\perp_{1K^*}$,$a^\perp_{2K^*}$ are determined in Ref.\cite{geg} and listed in the next section.

The strong interactions between the resonance and the final-state meson pair can be factorized into the time-like
form factor, which is guaranteed by the Watson theorem\cite{kw}. For the narrow resonances, the relativistic
Breit-Wigner (RBW)\cite{rbw} function is
a convenient model to well separate from any other resonant or nonresonant contributions with
the same spin, and has been widely used in the experimental data analyses. Here, the time-like form factor $F^{^\parallel}_{K\pi}(\omega^2)$ is parameterized with the RBW line shape and can be expressed as the following form\cite{factor1,factor2}
\be{}
F^\parallel_{K\pi}(\omega^2) = \frac{m_{K^*}^{2}}{m_{K^*}^{2}-\omega^2-im_{K^*}\Gamma_{K^*}(\omega^2)},
\en{}
where the $m_{K^*}$ and $\Gamma_{K^*}(\omega^2)$ are the pole mass and width. The mass dependent width $\Gamma_{K^*}(\omega^2)$ is define as
\be
\Gamma_{K^*}(\omega^2) = \Gamma_{K^*}\left(\frac{m_{K^*}}{\omega}\right)\left(\frac{|\overrightarrow{P_1}|}
{|\overrightarrow{P_0}|}\right)^{(2L_R+1)}\frac{1+(|\overrightarrow{P_0}|r_{BW})^2}{1+(|\overrightarrow{P_1}|r_{BW})^2},
\en
where $|\overrightarrow{P_1}|$ is the magnitude of the $K (\pi)$ momentum measured in the resonance $K^*$ rest frame,
while $|\overrightarrow{P_0}|$ is the value of $|\overrightarrow{P_1}|$ corresponding to $\omega = m_{K^*}$. $L_R$ is the orbital
angular momentum in the $K\pi$ system and $L_R = 1$ corresponds to the P-wave resonances.
Due to the limited studies on the form factor $F^\perp_{K\pi}(\omega^2)$, we use the two decay constants
$f_{K^*}^T$ and $f_{K^*}$ of the intermediate particle to determine it through the ratio $F^\perp_{K\pi}(\omega^2)/F^\parallel_{K\pi}(\omega^2)
\approx (f_{K^*}^T/f_{K^*})$\cite{pqcd3}.

For the wave function of the heavy $B_{(s)}$ meson\cite{bdas}, we take
\be\label{eq8}
\Phi_{B_{(s)}}(x, b) = \frac{1}{\sqrt{2 N_{c}}}\left(P\hspace{-2.5truemm}/_{B_{(s)}}+m_{B_{(s)}}\right) \gamma_{5} \phi_{B_{(s)}}(x, b).
\en
Here only the contribution of Lorentz structure $\phi_{B_{(s)}}(x, b)$ is taken into account, since the contribution
of the second Lorentz structure $\bar\phi_{B_{(s)}}$ is numerically small and has been neglected \cite{cdlu}. For the distribution amplitude $\phi_{B_{(s)}}(x, b)$ in Eq.\eqref{eq8}, we adopt the following
model
\be
\phi_{B_{(s)}}(x,b) = N_{B_{(s)}}x^2(1-x)^2exp\left(-\frac{x^2m_{B_{(s)}}^2}{2\omega_b^2}-\frac{\omega_b^2b^2}{2}\right),
\en
where the shape parameter $\omega_b=0.40\pm0.04(\omega_b=0.50\pm0.05)$ GeV  has been well fixed
by using the rich experimental data on the $B_{(s)}$ meson in many works, and the coefficient $N_{B_{(s)}}$ is determined by the normalization$\int_0^1 dx \phi_{B_{(s)}}(x,b=0) = 1$.
\subsection{Analytic formulae}

For the quasi-two-body decays $B\to K^*\gamma\to K\pi\gamma$, the effective Hamiltonian relevant to the $b\to s$ transition is given by\cite{heff}
\be
\bea
H_{eff} = &\frac{G_F}{\sqrt{2}}[\sum_{q=u,c}V_{qb}V_{qs}^*\{C_1(\mu)O_1^{(q)}(\mu)+C_2(\mu)O_2^{(q)}(\mu)\}\\
&-\sum_{i=3\sim8g}V_{tb}V_{ts}^*C_i(\mu)O_i(\mu)]+H.c.,
\ena
\en
where the Fermi coupling constant $G_F\simeq 1.166\times 10^{-5}GeV^{-2}$\cite{pdg}, $V_{qb}V_{qs}^*$ and $V_{tb}V_{ts}^*$
are the products of the CKM matrix elements. The scale $\mu$ separates the effective Hamiltonian into two distinct parts: the Wilson
coefficients $C_i$ and the local four-quark operators $O_i$. The local four-quark operators are written as
\begin{figure}[htbp]
\centering
\begin{minipage}[t]{0.45\linewidth}
\includegraphics[scale=0.4]{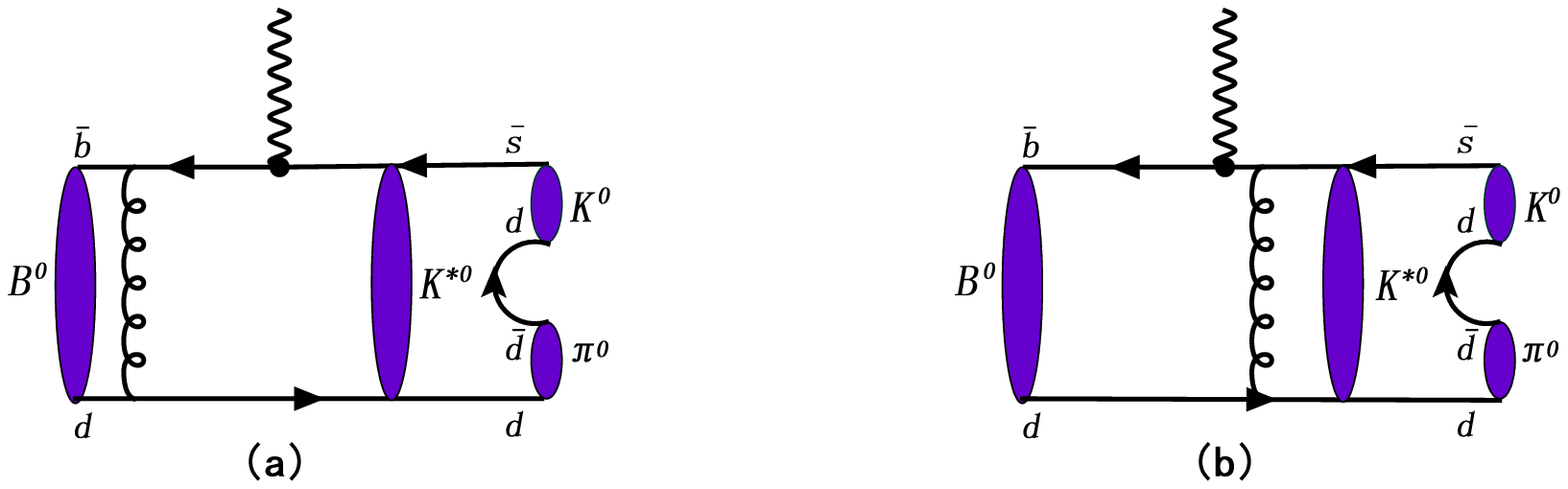}
\caption{Feynman diagrams from the operator $O_{7\gamma}$.}\label{img1}
\end{minipage}
\begin{minipage}[t]{0.45\linewidth}
\includegraphics[scale=0.4]{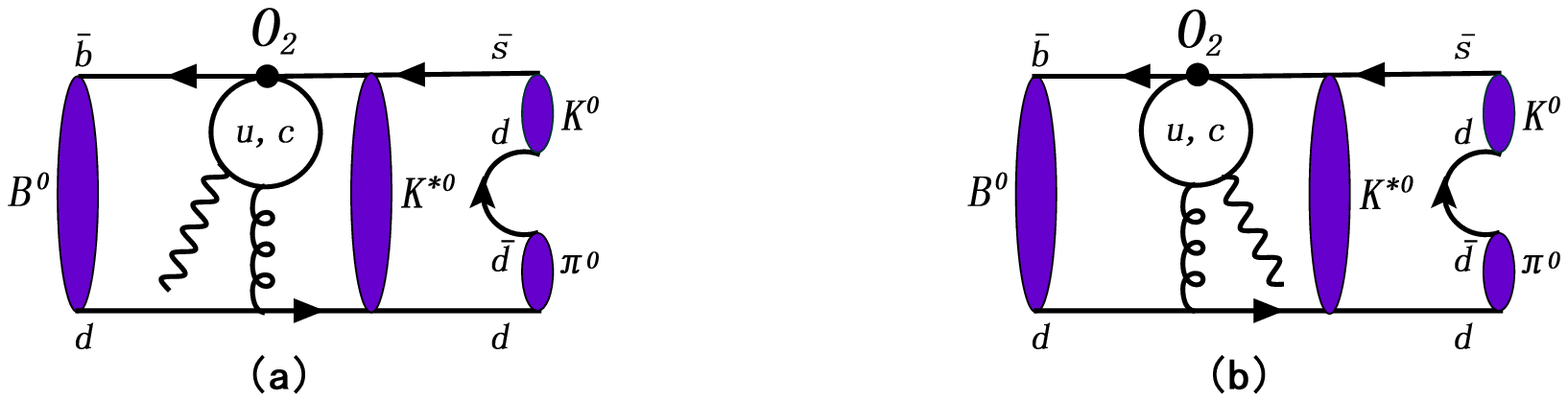}\caption{Quark-loop diagrams from the operator $O_2$ with the photon being emitted from the quark
loop.}\label{img3}
\end{minipage}
\end{figure}
\be
\bea
O_1^{(q)}&=(\Bar{s_i}q_j)_{V-A}(\Bar{q_j}b_i)_{V-A},O_2^{(q)}=(\Bar{s_i}q_i)_{V-A}(\Bar{q_j}b_j)_{V-A},O_3=(\Bar{s_i}b_i)_{V-A}\sum_q(\Bar{q_j}q_j)_{V-A},\\
O_4&=(\Bar{s_i}b_j)_{V-A}\sum_q(\Bar{q_j}q_i)_{V-A},O_5=(\Bar{s_i}b_i)_{V-A}\sum_q(\Bar{q_j}q_j)_{V+A,},O_6=(\Bar{s_i}b_j)_{V-A}\sum_q(\Bar{q_j}q_i)_{V+A},\\
O_{7\gamma}&=\frac{e}{8\pi^2}m_b\Bar{s}_i\sigma^{\mu\nu}(1+\gamma_5)b_iF_{\mu\nu},O_{8g}=\frac{g}{8\pi^2}m_b\Bar{s}_i\sigma^{\mu\nu}(1+\gamma_5)T_{ij}^ab_jG_{\mu\nu}^a,
\ena
\en
with the color indices i and j. Here $V\pm A$ refer to the Lorentz structures $\gamma_\mu(1\pm \gamma_5)$.
It is noticed that the terms associated with the strange quark mass in the $O_{7\gamma}$ and $O_{8g}$ operators have been dropped.
\begin{figure}[htbp]
\centering
\begin{minipage}[t]{0.45\linewidth}
\hspace{2mm}
\includegraphics[scale=0.4]{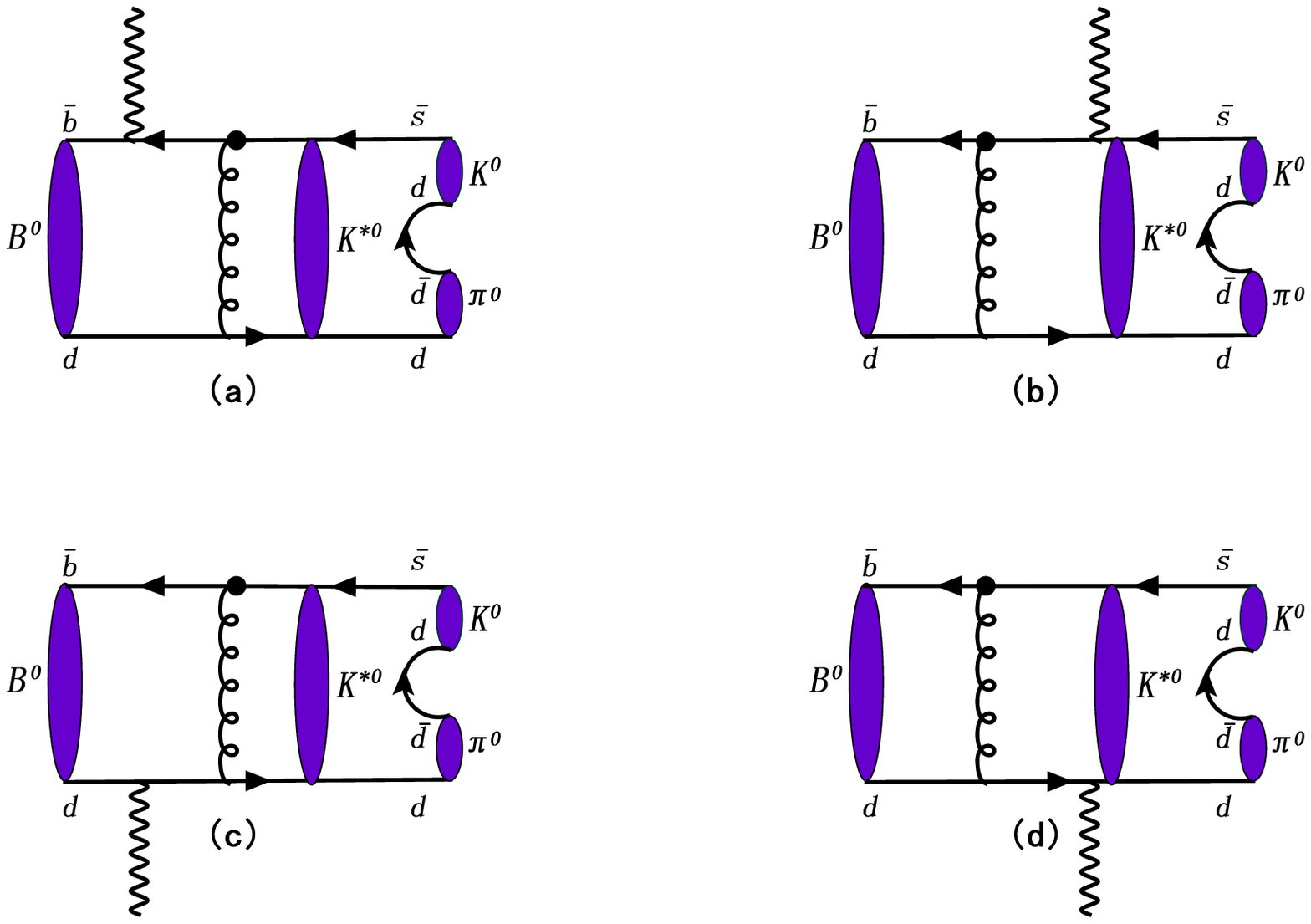}
\caption{Feynman diagrams from the operator $O_{8g}$.}\label{img2}
\end{minipage}
\begin{minipage}[t]{0.45\linewidth}
\hspace{2mm}
\includegraphics[scale=0.4]{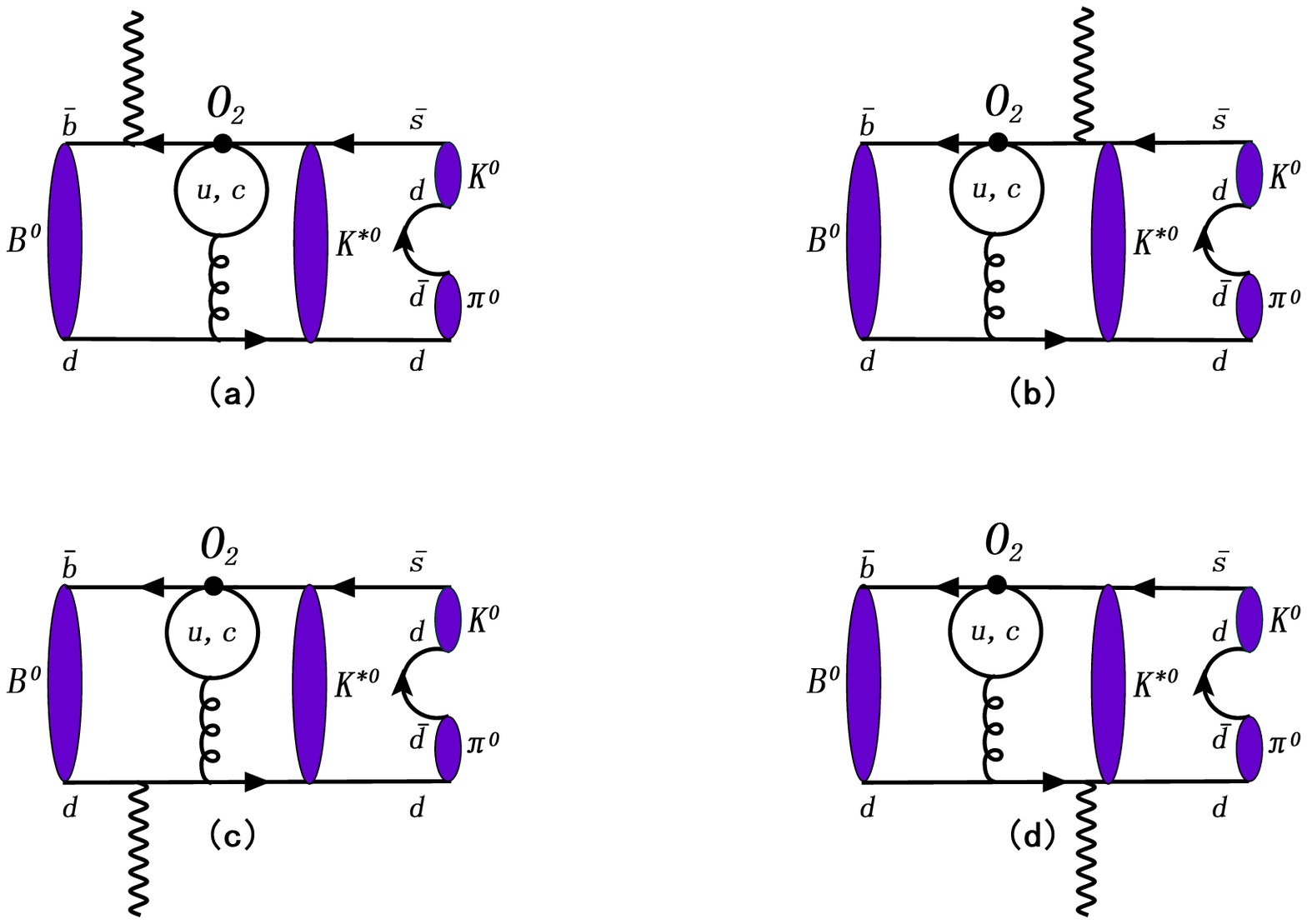}\caption{Quark-loop diagrams from the operator $O_2$ with a photon being emitted by an external
quark.}\label{img4}
\end{minipage}
\end{figure}
\begin{figure}[htbp]
\centering
\includegraphics[scale=0.4]{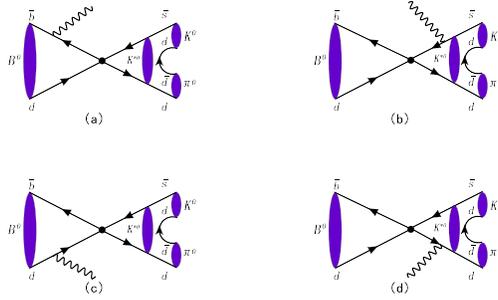}
\caption{Annihilation diagrams.}\label{img5}
\end{figure}

The typical Feynman diagrams at the leading order for the quasi-two-body
decays $B\to K^*\gamma\to K\pi\gamma$ (through $b\to s$ transition) is shown in Fig.\ref{img1}-\ref{img5}, where we take
$B\to K^{*0}\gamma\to K^0\pi^0\gamma$ decay as an example. The contributions from the $O_{7\gamma}, O_{8g}, O_2$ operators and the
annihilation type diagrams are involved, and the analysis formulas for the decay amplitudes of each Feynman diagram can be found in our previous work \cite{pqcd9}.
Certainly, the wave functions and the corresponding parameters need to be replaced in the calculations.

By combining the amplitudes from the different
Feynman diagrams, the total decay amplitude for the charged $B$ meson decay is given as
\be\label{eq12}
\bea
\mathcal{A}^i(B^+)=&\frac{G_F}{\sqrt{2}}V_{ub}^*V_{us}\left\{C_2\left(\mathcal{M}_{1u}^{i(a)}+\mathcal{M}_{1u}^{i(b)}(Q_u)+\mathcal{M}_{2u}^i\right)+a_1\left(\mathcal{M}_{ann}^{i(a,LL)}(Q_u)+\mathcal{M}_{ann}^{i(b,LL)}(Q_u)\right)\right\}\\
&+\frac{G_F}{\sqrt{2}}V_{cb}^*V_{cs}\left\{C_2\left(\mathcal{M}_{1c}^{i(a)}+\mathcal{M}_{1c}^{i(b)}(Q_u)+\mathcal{M}_{2c}^i\right)\right\}\\
&-\frac{G_F}{\sqrt{2}}V_{tb}^*V_{ts}\left\{C_{7\gamma}\mathcal{M}_{7\gamma}^i+C_{8g}\left(\mathcal{M}_{8g}^{i(a)}+\mathcal{M}_{8g}^{i(b)}(Q_u)\right)\right.\\
&+\left. (a_4+a_{10})\left(\mathcal{M}_{ann}^{i(a,LL)}(Q_u)+\mathcal{M}_{ann}^{i(b,LL)}(Q_u)\right)+(a_6+a_8)\mathcal{M}_{ann}^{i(SP)}(Q_u)\right\},
\ena
\en
where $i=R,L$ correspond to the contributions from the right-handed and left-handed photons, respectively, and the combinations of the Wilson coefficients are defined as
\be\label{eq11}
\bea
a_1 =& C_2+C_1/3, a_4 = C_4+C_3/3, a_6 = C_6+c_5/3,\\
a_8 =& C_8+C_7/3, a_{10} = C_{10}+C_9/3.
\ena
\en

Similarly, the total decay amplitudes for the decays $B_{(s)}^0\to K^{*}\gamma\to K\pi\gamma$ are listed as following
\be\label{eq13}
\bea
\mathcal{A}^i(B^0)=&\frac{G_F}{\sqrt{2}}V_{ub}^*V_{us}\left\{C_2\left(\mathcal{M}_{1u}^{i(a)}+\mathcal{M}_{1u}^{i(b)}(Q_d)+\mathcal{M}_{2u}^i\right)\right\}\\
&+\frac{G_F}{\sqrt{2}}V_{cb}^*V_{cs}\left\{C_2\left(\mathcal{M}_{1c}^{i(a)}+\mathcal{M}_{1c}^{i(b)}(Q_d)+\mathcal{M}_{2c}^i\right)\right\}\\
&-\frac{G_F}{\sqrt{2}}V_{tb}^*V_{ts}\left\{C_{7\gamma}\mathcal{M}_{7\gamma}^i+C_{8g}\left(\mathcal{M}_{8g}^{i(a)}+\mathcal{M}_{8g}^{i(b)}(Q_d)\right)\right.\\
&+(a_4-\frac{1}{2}a_{10})\left(\mathcal{M}_{ann}^{i(a,LL)}(Q_d)+\mathcal{M}_{ann}^{i(b,LL)}(Q_d)\right)+(a_6-\frac{1}{2}a_8)\left.\mathcal{M}_{ann}^{i(SP)}(Q_d)\right\},
\ena
\en
\be\label{eq14}
\bea
\mathcal{A}^i(B_s^0)=&\frac{G_F}{\sqrt{2}}V_{ub}^*V_{ud}\left\{C_2\left(\mathcal{M}_{1u}^{i(a)}+\mathcal{M}_{1u}^{i(b)}(Q_s)+\mathcal{M}_{2u}^i\right)\right\}\\
&+\frac{G_F}{\sqrt{2}}V_{cb}^*V_{cd}\left\{C_2\left(\mathcal{M}_{1c}^{i(a)}+\mathcal{M}_{1c}^{i(b)}(Q_s)+\mathcal{M}_{2c}^i\right)\right\}\\
&-\frac{G_F}{\sqrt{2}}V_{tb}^*V_{td}\left\{C_{7\gamma}\mathcal{M}_{7\gamma}^i+C_{8g}\left(\mathcal{M}_{8g}^{i(a)}+\mathcal{M}_{8g}^{i(b)}(Q_s)\right)\right.\\
&+(a_4-\frac{1}{2}a_{10})\left(\mathcal{M}_{ann}^{i(a,LL)}(Q_s)+\mathcal{M}_{ann}^{i(b,LL)}(Q_s)\right)+(a_6-\frac{1}{2}a_8)\left.\mathcal{M}_{ann}^{i(SP)}(Q_s)\right\}.
\ena
\en

Then the differential decay rate can be described as
\be\label{eq15}
\frac{d\mathcal{B}}{d\omega}=\tau_B\frac{|\overrightarrow{P_1}||\overrightarrow{P_3}|}{32\pi^3m_B^3}{\sum_{i=R,L}|\mathcal{A}^i|}^2,
\en
where the squared amplitudes for the $B_{(s)}$ meson decays are summed in the helicity basis, and $\tau_{B}$ is the mean lifetime of $B_{(s)}$ meson, the kinematic variables $|\overrightarrow{P_1}|$ and $|\overrightarrow{P_3}|$ denote the magnitudes of the $K^*$ and $\gamma$ momenta in the center-of-mass frame of the $K\pi$ pair,
\be
\bea
|\overrightarrow{P_1}|&=\frac{1}{2}\sqrt{[(w^2-(m_K+m_{\pi})^2)(w^2-(m_K-m_{\pi})^2)]/w^2},\\
|\overrightarrow{P_3}|&=\frac{1}{2}(m_B^2-w^2)/m_B.
\ena
\en
{\centering\section{NUMERICAL RESULTS}\label{results}}

The adopted input parameters in our numerical calculations are summarized as following (the masses,
decay constants and QCD scale are in units of GeV, the B meson lifetimes are in units of ps)\cite{canshu1,canshu2,pdg}
\be
\bea
\Lambda_{QCD} &= 0.25, m_{B^+} = 5.279, m_{B^0} = 5.280, m_{B_s} = 5.267,\\
m_b &= 4.8, m_{K^{\pm}} = 0.494, m_{K^0} = 0.498, m_{\pi^{\pm}} = 0.140,\\
m_{\pi^0} &= 0.135, m_{K^{*0}} = 0.89555, m_{K^{*\pm}} = 0.89176,  f_B = 0.19,\\
f_{B_s} &= 0.23, \tau_{B^+} = 1.638,\tau_{B^0} = 1.520, \tau_{B_s^0} = 1.509,\\
\Gamma_{K^{*0}} &= 47.3, \Gamma_{K^{*\pm}} = 50.3, r_{BW} = 4GeV^{-1}, f_{K^*} = 0.217,\\
f_{K^*}^T &= 0.185, a^\perp_{1K^*} = 0.31\pm0.16, a^\perp_{2K^*} = 1.188\pm0.098.
\ena
\en
As to the Cabibbo-Kobayashi-Maskawa (CKM) matrix elements, we employ the Wolfenstein parametrization with the inputs\cite{pdg}
\be
\bea
\lambda&=0.22453 \pm 0.00044, \quad A=0.836 \pm 0.015,\\
\bar{\rho}&=0.122_{-0.017}^{+0.018}, \quad \bar{\eta}=0.355_{-0.011}^{+0.012}.
\ena
\en

By using the differential branching ratio in Eq.\eqref{eq15}
and the squared amplitudes in Eq.\eqref{eq12}, Eq.\eqref{eq13}, integrating over the full $K\pi$ invariant mass region
$(m_K + m_{\pi})\leq \omega\leq M_{B_{(s)}}$ for the resonant components, we obtain the branching ratios for these quasi-two-body decays as
\be\label{bran}
\bea
Br\left(B^0\to K^{*0} \gamma \to K^+\pi^- \gamma \right) &= \left({3.08}_{-0.86-0.30-0.36-0.08}^{+1.29+0.33+0.24+0.12}\right)\times{10}^{-5},\\
Br\left(B^0\to K^{*0} \gamma \to K^0\pi^0 \gamma \right) &= \left({1.55}_{-0.44-0.15-0.19-0.04}^{+0.64+0.16+0.12+0.06}\right)\times{10}^{-5},\\
Br\left(B^+\to K^{*+} \gamma \to K^+\pi^0 \gamma \right) &= \left({1.72}_{-0.46-0.17-0.12-0.07}^{+0.67+0.20+0.08+0.10}\right)\times{10}^{-5},\\
Br\left(B^+\to K^{*+} \gamma \to K^0\pi^+ \gamma \right) &= \left({3.21}_{-0.87-0.31-0.23-0.12}^{+1.25+0.36+0.20+0.15}\right)\times{10}^{-5},
\ena
\en
where the first source of errors originates from the shape parameter of the B meson DA, $\omega_B=0.4\pm0.04$ GeV,
the second error is from the Gegenbauer coefficients in the kaon-pion distribution amplitudes
:$a^{\perp}_{1K^*}=0.31\pm0.16,a^{\perp}_{2K^*}=1.188\pm0.098$, the last two errors are induced by
the next-to-leading-order effects in PQCD approach: changing the hard scale t from 0.80t to 1.2t and
the QCD scale $\Lambda_{QCD}=0.25\pm0.05$GeV, respectively.
From our results, one can see that the dominant theoretical error comes from the uncertainty of $\omega_B$,
which is close to $40\%$. The error induced by the Gegenbauer coefficients in $K\pi$ pair distribution amplitudes is smaller
and about $10\%$.
 These four decays are mediated by $b\to s$
transition, which is proportional to $V_{tb}V^*_{ts}\sim \lambda^2$. In our calculations, the Feynman diagrams
from the operator $O_{7\gamma}$ give the dominant contributions.

If assuming the isospin conservation for the strong
decays $K^*\to K\pi$, one can obtain the following relations
\be\label{22}
\bea
\frac{\Gamma(K^{*0}\to K^+\pi^-)}{\Gamma(K^{*0}\to K\pi)}=2/3,\;\;\frac{\Gamma(K^{*0}\to K^0\pi^0)}{\Gamma(K^{*0}\to K\pi)}=1/3,\\
\frac{\Gamma(K^{*+}\to K^0\pi^+)}{\Gamma(K^{*+}\to K\pi)}=2/3,\;\;\frac{\Gamma(K^{*+}\to K^+\pi^0)}{\Gamma(K^{*+}\to K\pi)}=1/3.
\ena
\en
Under the narrow width approximation, the branching ratios of these quasi-two-body decays can be expressed as
\be\label{21}
\bea
Br\left(B^0\to K^{*0} \gamma \to K^+\pi^- \gamma \right) &= Br(B^0\to K^{*0}\gamma)\cdot Br(K^{*0}\to K^+\pi^-),\\
Br\left(B^0\to K^{*0} \gamma \to K^0\pi^0 \gamma \right) &= Br(B^0\to K^{*0}\gamma)\cdot Br(K^{*0}\to K^0\pi^0),\\
Br\left(B^+\to K^{*+} \gamma \to K^+\pi^0 \gamma \right) &= Br(B^+\to K^{*+}\gamma)\cdot Br(K^{*+}\to K^+\pi^0),\\
Br\left(B^+\to K^{*+} \gamma \to K^0\pi^+ \gamma \right) &= Br(B^+\to K^{*+}\gamma)\cdot Br(K^{*+}\to K^0\pi^+).
\ena
\en
Using the experimental data given in Eq.\eqref{23}, which were measured by Belle II in the last year,
combining with isospin conservation Eq.\eqref{22} and narrow width
approximation Eq.\eqref{21}, we can estimate the branching ratios of the quasi-two-body decays as following
\be
\bea
Br\left(B^0\to K^{*0} \gamma \to K^+\pi^- \gamma \right) &= \left(3.00 \pm 0.20 \pm 0.13 \right)\times 10^{-5},\\
Br\left(B^0\to K^{*0} \gamma \to K^0\pi^0 \gamma \right) &= \left(1.47 \pm 0.30 \pm 0.20 \right) \times 10^{-5},\\
Br\left(B^+\to K^{*+} \gamma \to K^+\pi^0 \gamma \right) &= \left(1.67 \pm 0.17 \pm 0.13 \right) \times 10^{-5},\\
Br\left(B^+\to K^{*+} \gamma \to K^0\pi^+ \gamma \right) &= \left(3.60 \pm 0.40 \pm 0.27 \right) \times 10^{-5}.
\ena
\en
One can find that these estimates and our predictions are consistent well with each other,
so it is reasonable to extend  the PQCD approach to
the B meson quasi-two-body decays.

From the numerical results as given in Eq.\eqref{bran}, we calculate the relative ratio $R_{1,2}$ between the branching ratios of the charged and neutral
B meson decays
\be
\bea
R_1&=&\frac{Br\left(B^+\to K^{*+} \gamma \to K^0\pi^+ \gamma \right)}{Br\left(B^0\to K^{*0} \gamma \to K^+\pi^- \gamma \right)}=1.04^{+0.63}_{-0.46},\\
R_2&=&\frac{Br\left(B^+\to K^{*+} \gamma \to K^+\pi^0 \gamma\right)}{Br\left(B^0\to K^{*0} \gamma \to K^0\pi^0 \gamma\right)}=1.11^{+0.67}_{-0.49}.
\ena
\en
If we assume the branching ratio of
the decay $K^{*0}\to K\pi$ to be $100\%$, the isospin conservation
and the narrow width approximation to be ture, we can relate them with the ratio $R=\frac{Br\left(B^+\to K^{*+}
\gamma\right)}{Br\left(B^0\to K^{*0} \gamma\right)}$.  Using the data
from PDG \cite{pdg}, one can get the ratio is $R=\frac{Br\left(B^+\to K^{*+}
\gamma\right)}{Br\left(B^0\to K^{*0} \gamma\right)}=0.94\pm0.08$. Using
the update data measured by Belle II \cite{cankao}, one can get the ratio is
$R=\frac{Br\left(B^+\to K^{*+} \gamma\right)}{Br\left(B^0\to K^{*0}
\gamma\right)}=1.16\pm0.14$. The values of these ratios
once again support the usability and rationality of the PQCD factorization for the B meson quasi-two-body decays.

The branching ratios of the two-body decays $B^{+,0}\to K^{*+,0} \gamma$ have been calculated in PQCD approach \cite{matsumori}, where the results were given as
$Br(B^{0}\to K^{*0} \gamma)=(5.8\pm2.9)\times10^{-5}, Br(B^{+}\to K^{*+} \gamma)=(6.0\pm3.0)\times10^{-5}$. Two years later, they were
updated with $Br(B^{0}\to K^{*0} \gamma)=(3.81^{+1.73+0.55+0.11}_{-1.27-0.38-0.11})\times10^{-5},
Br(B^{+}\to K^{*+} \gamma)=(3.58^{+1.76+0.54+0.11}_{-1.28-0.40-0.11})\times10^{-5}$ \cite{wangw}.
Compared with these two group
calculations, there still exists appear differences. Compared all these theoretical results with the new update data
measured by Belle II, it supports that studying these B meson quasi-two-body decays is more
appropriate in the
three-body framework than in the two-body one. Under the times of high precision measurement, these results should be
further tested carefully in the
LHCb experiments.

By using the same two-meson DAs for the $K\pi$, we also calculate the branching ratios for the decays
$B_s^0\to \bar{K}^{*0}\gamma\to K^0\pi^0(K^-\pi^+ )\gamma$ and obtain the results as
\be\label{Bsdecay}
\bea
Br\left(B_s^0\to \bar{K}^{*0}\gamma \to K^0\pi^0 \gamma \right) &= \left({0.35}_{-0.11-0.03-0.02-0.02}^{+0.16+0.04+0.00+0.02}\right)\times{10}^{-6},\\
Br\left(B_s^0\to \bar{K}^{*0}\gamma \to K^-\pi^+ \gamma \right) &= \left({0.69}_{-0.21-0.07-0.03-0.03}^{+0.33+0.08+0.00+0.04}\right)\times{10}^{-6}.
\ena
\en
These two decays are induced by the $b\to d$ transition, which is proportional to $V_{tb}V_{td}^*\sim \lambda^3$
and expected to be suppressed by one order of magnitude relative to those induced by the $b\to s$ transition. From Eq.\eqref{Bsdecay}, one can get the
branching ratio of the two-body decay $B_s\to\bar{K}^{*0}\gamma$ is $(1.04^{+0.51}_{-0.34})\times10^{-6}$ through $\bar{K}^{*0}\to K^-\pi^+$ internal decay mode
and $(1.05^{+0.50}_{-0.35})\times10^{-6}$ through $\bar{K}^{*0}\to K^0\pi^0$.
Our prediction is consistent well with the result given in Ref.\cite{wangw}
\be
Br\left(B_s^0\to \bar K^{*0}\gamma\right) =\left(1.11_{-0.32-0.12-0.07}^{+0.42+0.15+0.16}\right)\times{10}^{-6}.
\en
The consistency of these results indicates that the PQCD approach can be applicable to the two-body and three-body decays at the same time.
\begin{figure}[htbp]
\begin{center}
\includegraphics[width=0.7\textwidth]{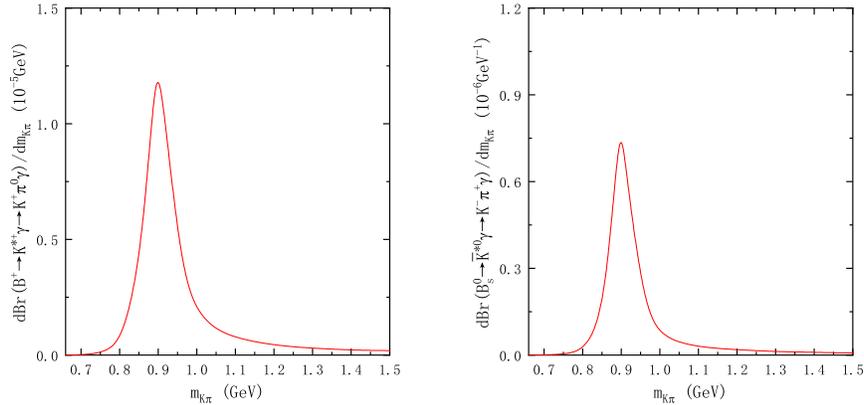}
\caption{The predicted $B^+\to K^{*+}\gamma\to K^+\pi^0\gamma$ (left) and $B_{s}\to \bar K^{*0}\gamma\to K^-\pi^+\gamma$ (right) decay spectra in the $K\pi$ invariant mass.}\label{img6}
\end{center}
\end{figure}

We also predict the $\omega-$dependences of $B^+\to K^{*+}\gamma\to K^+\pi^0\gamma$ and $B_{s}\to \bar K^{*0}\gamma\to K^-\pi^+\gamma$
decay spectra shown in Fig.\ref{img6}, which exhibit a maximum at
the $K\pi$ invariant mass around 0.895 GeV. The curves for the other three B meson decay modes are similar, since the same time-like form factors for the $K\pi$ DAs.
It is easy to see that the main contribution to the branching ratio
comes from the region around the pole mass of the $K^*$ resonance as we expected. For example, the central values of the
branching ratio $BR(B^+\to K^{*+}\gamma\to K^+\pi^0\gamma)$ are $0.89\times10^{-5}$ and $1.30\times10^{-5}$ when
we integrate over $\omega$ by limiting the ranges of $\omega=[m_{K^*}-0.5\Gamma_{K^*},m_{K^*}+0.5\Gamma_{K^*}]$
and $\omega=[m_{K^*}-\Gamma_{K^*},m_{K^*}+\Gamma_{K^*}]$, repectively, which amount to $52\%$ and $76\%$ of the total branching ratio
$BR(B^+\to K^{*+}\gamma\to K^+\pi^0\gamma)=1.72\times10^{-5}$.

The direct CP asymmetry of the $B_{(s)} \rightarrow K^{*} \gamma\to K\pi\gamma$ is defined by
\be
A_{C P}=\frac{\Gamma\left(\bar{B}_{(s)} \rightarrow \bar{K}^{*} \gamma\to \overline{(K\pi)}\gamma\right)-\Gamma\left(B_{(s)} \rightarrow K^{*}\gamma\to K\pi\gamma\right)}
{\Gamma\left(\bar{B}_{(s)} \rightarrow \bar{K}^{*} \gamma\to \overline{(K\pi)}\gamma\right)+\Gamma\left(B_{(s)} \rightarrow K^{*} \gamma\to K\pi\gamma\right)}
\en
For our considered decays, it is mainly induced by the interference between the contributions from $O_{7\gamma}$ operator and
the tree operator which is proportional to $V^*_{ub}V_{us(d)}$. We predict the direct CP asymmetries as
\be
\bea
A_{C P}\left(B^0\to K^{*0} \gamma \to K^+\pi^- \gamma \right) &= \left({-0.58}_{-0.00-0.65-0.09-0.21}^{+0.04+0.61+0.06+0.00}\right)\times{10}^{-2},\\
A_{C P}\left(B^0\to K^{*0} \gamma \to K^0\pi^0 \gamma \right) &= \left({-0.55}_{-0.00-0.77-0.00-0.08}^{+0.37+0.68+0.43+0.03}\right)\times{10}^{-2},\\
A_{C P}\left(B^+\to K^{*+} \gamma \to K^+\pi^0 \gamma \right) &= \left({-0.79}_{-0.21-0.74-0.00-0.15}^{+0.24+0.37+0.79+0.18}\right)\times{10}^{-2},\\
A_{C P}\left(B^+\to K^{*+} \gamma \to K^0\pi^+ \gamma \right) &= \left({-0.39}_{-0.16-0.80-0.34-0.00}^{+0.35+0.70+0.02+0.46}\right)\times{10}^{-2},\\
A_{C P}\left(B_s^0\to \bar{K}^{*0} \gamma \to K^0\pi^0 \gamma \right) &= \left({4.22}_{-0.00-0.95-0.05-0.00}^{+0.83+0.70+0.65+0.10}\right)\times{10}^{-2},\\
A_{C P}\left(B_s^0\to \bar{K}^{*0} \gamma \to K^-\pi^+ \gamma \right) &= \left({4.17}_{-0.21-0.84-0.42-0.11}^{+0.00+0.52+0.00+0.00}\right)\times{10}^{-2}.
\ena
\en
Where we can find that the direct CP violations of the $B^{0,+}$ decays induced by $b\to s$ transition are less than $1\%$.
It is consistent with those of the two-body decays $B^{+,0} \rightarrow K^{*+,0} \gamma$ predicted by PQCD approach \cite{matsumori}
$A_{CP}(B^+\to K^{*+}\gamma)=-(0.57\pm0.43)\times10^{-2}$ and $A_{CP}(B^0\to K^{*0}\gamma)=-(0.61\pm0.46)\times10^{-2}$, and
they were recalculated
to be $A_{CP}(B^+\to K^{*+}\gamma)=-(0.40\pm0.43)\times10^{-2}$ and $A_{CP}(B^0\to K^{*0}\gamma)=-(0.30\pm0.00)\times10^{-3}$ in Ref.\cite{wangw}.
On the experimental side, the direct CP asymmetries of the decays $B^{+,0} \rightarrow K^{*+,0}\gamma$ are given in PDG \cite{pdg}
\be
\bea
A_{CP}\left(B^{+} \rightarrow K^{*+}\gamma\right) &= (1.4\pm 1.8)\%,\\
A_{CP}\left(B^{0} \rightarrow K^{*0} \gamma\right) &= -(0.6\pm 1.1)\%,
\ena
\en
where $A_{CP}\left(B^{+} \rightarrow K^{*+}\gamma\right)$ is larger than $1\%$ and contrary in sign with the PQCD predictions. Certainly, there still exists
larger errors. We hope that this
divergence can be clarified by the future LHCb and SuperKEKB experiments through measuring the three-body decays
$B^+\to K^{*+} \gamma \to K^+\pi^0 \gamma$ and $B^+\to K^{*+} \gamma \to K^0\pi^+ \gamma$. If more than a few percent value of the direct
CP violation is confirmed in the furture, one can consider that some new physics might contribute to these channels. While the direct CP violations for the $B_s$
decays induced by the $b\to d$ transition are much larger than those of the $B^{+,0}$ decays, it is because that the product of the CKM matrix element for electro-magnetic
penguin operator is $V^*_{tb}V_{td}\sim\lambda^3$, and that for the tree operator is either proportional to $V^*_{cb}V_{cd}\sim\lambda^3$
or $V^*_{ub}V_{ud}\sim\lambda^3$. That is to say the tree contribution is not suppressed and can be comparative with the penguin contribution. As we know
the direct CP violation arises from the interference between the tree and penguin contributions. So one can expect the relatively large CP
asymmetries for these two $B_s$ decays. In the previous PQCD calculations \cite{wangw}, the authors obtained the results as
\be
\bea
A_{CP}\left(\bar{B}_{s}^{0} \rightarrow K^{* 0} \gamma\right) =\left(12.7_{-0.5-2.3-0.9}^{+0.1+1.6+0.5}\right)\times{10}^{-2},
\ena
\en
which is indeed much larger than those of $B^{0,+}$ decays. Though this prediction is larger than our predictions,
it is clear that the direct CP violation for the decay $B_{s}^{0} \rightarrow \bar K^{*0} \gamma$  has a positive sign,
which is contrary to those of the $B^{0,+} \rightarrow K^{*0,+}\gamma$ decays.

{\centering\section{SUMMARY}\label{suma}}

In this work, we analyzed the three-body radiative decays $B_{(s)}\to K^*\gamma\to K\pi\gamma$ with
$K\pi$ pair originating from the intermediate state $K^*$ by using the PQCD approach. Under the quasi-two-body-decay mechanism, the $K\pi$ pair
distribution amplitudes (DAs) are introduced, which include the final-state interactions between the $K\pi$ pair in the resonant region.
Both the resonant and nonresonant contributions are described by the time-like form factor
$F_{K\pi}$, which are parameterized by using the relativistic Breit-Wigner formula for the P-wave resonance $K^*$.
Under the condition of the narrow width approximation and the isospin conservation,  the branching ratios for
the decays $B_{(s)}\to K^*\gamma\to K\pi\gamma$ are consistent with those of the two-body decays $B_{(s)}\to K^*\gamma$ calculated by
the previous PQCD approach, which verified that the PQCD approach can be extended to B meson three-body decays. What is more important, our
predictions are much closer to the data recently measured by Belle II. It indicates that studying the B meson quasi-two-body decays is more
appropriate in the
three-body framework than in the two-body one. For the decays $B_{u,d}\to K^{*}\gamma\to K\pi \gamma$ induced by $b\to s$ transition,
their
direct CP violations are small and less than $1\%$. If more than a few percent value of the direct
CP violation is confirmed in the future, we can consider that some new physics might contribute to these channels. For the decays
$B_s\to K^{*}\gamma\to K\pi\gamma$ induced by  $b\to d$ transition, there exists stronger interference between the tree
and the penguin contributions, so relatively large CP asymmetries can be observed, which can be tested in the LHCb and Belle II experiments.

{\centering\section*{Acknowledgment}}

We thank Prof. Hsiang-nan Li for valuable discussion. This work is partly supported by the National Natural Science
Foundation of China under Grant No. 11347030, by the Program of
Science and Technology Innovation Talents in Universities of Henan
Province 14HASTIT037.

\end{document}